\documentclass[twocolumn,showpacs,preprintnumbers,amsmath,prl,amssymb,epsf]{revtex4}
\usepackage{graphicx}
\usepackage[dvips]{color}
\begin{document}
\title{Measurement of Photon Statistics with Live Photoreceptor Cells}
\author{Nigel Sim$^{1}$, Mei Fun Cheng$^{1}$, Dmitri Bessarab$^{2}$, C. Michael Jones$^{2}$, and Leonid A. Krivitsky$^{1,}$}
\email{Leonid_Krivitskiy@dsi.a-star.edu.sg}
\affiliation{$^1$ Data Storage Institute, Agency for Science Technology and Research (A-STAR), 117608 Singapore\\
$^2$ Institute of Medical Biology, Agency for Science Technology and Research (A-STAR), 138648 Singapore} \vskip 24pt

\begin{abstract}
\begin{center}\parbox{14.5cm}
{We analyzed the electrophysiological response of an isolated rod photoreceptor of \textit{Xenopus laevis} under stimulation by coherent and pseudo-thermal light sources. Using the suction electrode technique for single cell recordings and a fiber optics setup for light delivery allowed measurements of the major statistical characteristics of the rod response. The results indicate differences in average responses of rod cells to coherent and pseudo-thermal light of the same intensity and also differences in signal-to-noise ratios and second order intensity correlation functions. These findings should be relevant for interdisciplinary studies seeking applications of quantum optics in biology.}
\end{center}
\end{abstract}
\pacs{42.50.Ar, 42.66.Lc,	87.80.Jg}
 \maketitle \narrowtext
\vspace{-10mm}

Eyes of living organisms represent advanced light harvesting systems, developed through hundreds of millions years of evolution. Some of their features are comparable or even superior to existing man-made photodetection devices. For example, rod photoreceptor cells of the retina, which are responsible for night vision and form the focus of the present study, represent  miniaturized photodetectors containing a photosensitive element (rhodopsin pigment) along with a "built-in" chemical power supply (ATP produced by mitochondria) \cite{Dowling}. They have sensitivity down to single-photon level, and demonstrate a remarkable low-noise operation \cite{Rieke}. Understanding such properties of nature-given photodetectors stimulates considerable interest in interfacing them with sources of non-classical light, such as light with a "fixed" number of photons, and "squeezed" light \cite{DNK}. Implementation of such "bio-quantum" interfaces would contribute to a better understanding of the performance of the visual system near the threshold \cite{Teichmodulation}, allow a reference-free calibration of its quantum efficiency \cite{calibration}, along with a direct detection of entangled multi-photon states \cite{Gisin}. Study of the statistics of rod  responses to excitations by classical light sources allows careful characterization of rods, and open possibilities for future studies of non-classical light.

Vertebrate retinal photoreceptor rod cells, further referred to as rods, are densely packed in the retina laying the inner surface of an eyeball \cite{Dowling}. Individual rods consist of two distinctive functional regions: extended rod-like outer segment (OS) which is filled with rhodopsin photopigment molecules, and shorter rounded inner segment (IS) which contains various components of cell machinery. In the dark, the electrochemical gradient across the rod cell membrane causes a continuous flux of ions ($Na^{+},K^{+},Ca^{2+}$) in and out of the cell through aqueous pores in the membrane, referred to as ion channels. A photon, impinging on the OS, causes isomerization of a rhodopsin molecule, which initiates a chain of intracellular reactions, referred to as  phototransduction cascade. A fraction of ion channels closes, thus resulting in hyperpolarisation of the cell. Illumination with a relatively bright flash closes all light sensitive channels in the OS, and bleaches the rod. A recovery process regulates re-opening of the ion channels to the resting state.
Accurate measurement of the ion current flux through the rod membrane is possible with the suction-electrode technique \cite{Bailorsuction}. In this case, the OS of the rod is drawn in a tight fitting glass microelectrode filled with the physiological solution. The rod functions in a way similar to \textsl{in-vivo} system, however the ion current through the membrane is now re-directed to a low noise amplifier.

In earlier experiments isolated rods were mainly stimulated by multi-mode thermal sources (lamps, LEDs), whose statistics, however, was not within the scope of interest of those works \cite{BaylorSingle,Lambnoise,riekebaylor,riekenoise}. Influence of well controlled light statistics on the response of the entire visual system was studied only in behavioral experiments with human subjects \cite{Teichmodulation,Teichmodulation1}. However, noise properties of individual rods are difficult to access in this case, as several hundred of rods are simultaneously illuminated, and their collective response undergoes several intermediate stages of the visual processing. In the present study, influence of controllable photon fluctuations on the response of an isolated rod is analyzed for the first time. Using well characterized coherent and pseudo-thermal light sources, together with a fiber optics setup for light delivery allows us to determine major statistical characteristics of the rod in the whole dynamic range of its response.

The probability distribution of photon numbers is given by \cite{MandelWolf}: 
\begin{equation}
P_{ph}(m) = \begin{cases} \frac{{\overline{m}}^{m}e^{-\overline{m}}}{m!}, & \mbox{for the coherent source}\\ 
\frac{\overline{m}^m}{(1+\overline{m})^{m+1}}, & \mbox{for the thermal source} \\ \end{cases}
\end{equation}
where $\overline{m}$ is the average number of photons. Based on earlier findings, it is further assumed that each photon isomerizes only a single rhodopsin molecule \cite{howphotons}. In this case the probability of isomerization of $n$ molecules by $m$ impinging photons is given by the binomial distribution: $P(n|m)={m\choose n}\eta^n(1-\eta)^{m-n}$, where $\eta$ is defined as an overall quantum efficiency of the photodetection process \cite{MandelWolf}. Then, the probability of isomerization of $n$ molecules by the fluctuating light source is given by: 
\begin{equation}
P_{I}(n) = \sum^{\infty}_{m=0}P(n|m)P_{ph}(m).
\end{equation}
It is further assumed, that in the absence of saturation subsequent electrophysiological responses to individual isomerizations are additive, and have a standard Gaussian shape \cite{BaylorSingle}. In this case, the probability of observing the amplitude of the photocurrent $A$ in response to $n$ isomerizations is given by:
\begin{equation}
\widetilde{P}(A|n) =\frac{1}{\sqrt{2\pi(\sigma^{2}_{D}+n\sigma^{2}_{A})}}\mbox{exp}{\left(-\frac{(A-n\overline{A_{0}})^{2}}{2(\sigma^2_{D}+n\sigma^{2}_{A})}\right)},
\end{equation}
where $\overline{A_{0}}$ is the average, and $\sigma_{A}$ is the standard deviation of the photocurrent amplitude in response to a single isomerization, and $\sigma_{D}$ is the standard deviation of the photocurrent in darkness. Hence the final probability of observing the photocurrent amplitude $A$ is given by averaging over $P_{I}(n)$: 
\begin{equation}
P(A) = \sum^{\infty}_{n=0}\widetilde{P}(A|n)P_{I}(n).
\end{equation}
In order to account for the saturation of the photocurrent, $P(A)$ is truncated at the saturation amplitude $A_{S}$, so that the probability of observing the amplitude $A>A_{S}$ just adds up and results in observing $A_{S}$:
\begin{equation}
P_{S}(A) = \begin{cases} P(A), & \mbox{if } A<A_{S}\\ 
1-\int^{A_{S}}_{A=0}P(A)dA, & \mbox{if } A=A_{S} \\ 0, & \mbox{if } A>A_{S}\end{cases}
\end{equation}
Eq. (5) allows calculation of any statistical moment of the photocurrent amplitude, given by $\overline{A^{k}}=\int^{\infty}_{A=0}A^{k}P_{S}(A)dA$. The relation between the average $\overline{A}$ and the variance $\mbox{Var}A\equiv\overline{A^{2}}-\overline{A}^{2}$, can be conveniently characterized by a dimensionless signal-to-noise ratio $\mbox{SNR}\equiv \overline{A}/\sqrt{\mbox{Var}(A)}$.

Investigation of joint statistics of the rod response with some reference detector, measured in a Hanbury-Brown and Twiss (HBT) type experiment \cite{HBT, Klyshkophysel}, allows finding the second order intensity correlation function of the light field $g^{(2)}$. The  value of $g^{(2)}$ 
may evidence non-classical properties of light \cite{MandelWolf}. From the joint pulse-to-pulse data of photocurrent amplitudes of the rod and photocounts of a conventional avalanche photodiode (APD), $g^{(2)}$ is found as:
\begin{equation}
g^{(2)}\equiv\frac{\overline{AK}}{\overline{A}\mbox{ }\overline{K}}
=\frac{\int^{\infty}_{A=0}\sum^{\infty}_{K=0}AKP(K)P_{S}(A)dA}{\overline{A}\mbox{ }\overline{K}},
\end{equation}
where $\overline{K}=\eta_{D}\overline{m}$ is the average number of APD photocounts, $\eta_{D}$ is the APD quantum efficiency, and $P(K)$ is the probability distribution of APD photocounts, which is similar to (1).

In the experiment (see Fig.1a) the beam of a 50 mW continuous wave frequency-doubled Nd:YAG laser at 532 nm (Photop, DPL-2050) was chopped by an optical shutter (Melles Griot) in 30 ms duration pulses with a repetition rate of 10-15 seconds. The pseudo-thermal source, shown in the inset of Fig.1a, was realized by focusing the beam of the same laser with a lens (\textit{f}=100 mm) into a rotating ground glass disc (GD, 1500 grit), followed by a 0.5 mm iris diaphragm (D), placed at a distance 1200 mm \cite{thermalsourcedisk}. The diameter of the diaphragm was set to be much smaller than the size of a single scattered speckle. Then, either the laser beam (for the experiment with the coherent source) or a speckle (for the experiment with the pseudo-thermal source) was attenuated by a variable neutral density filter (NDF), and directed to a HBT setup, consisting of a non-polarizing 50/50 beamsplitter (BS) and two single mode optical fibers (SM 460-HP) with equalized coupling efficiencies preceded by aspheric lenses  (\textit{f}=6.24 mm). One of the fibers was used to stimulate the cell, whilst the other one was connected to an actively quenched APD (Perkin Elmer, SPCM-AQRH-15-FC). The number of photons used to stimulate the cell was calculated from the number of APD photocounts in each pulse, assuming the APD quantum efficiency $\eta_{D}=$50\%~\cite{APD}. The statistics of coherent and pseudo-thermal sources are checked in independent measurements by the APD (see supplementary materials). 

\begin{figure}
\includegraphics[width=8.0cm]{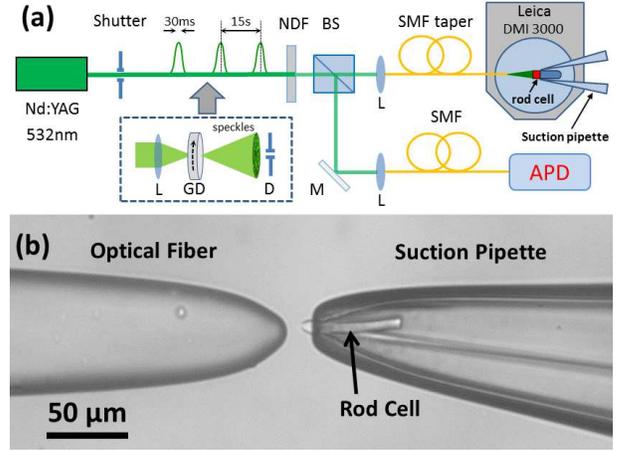}
\caption{(color online) (a) Optical layout. A laser beam is chopped by a shutter, attenuated by a neutral density filter (NDF), and split by a 50/50 beamsplitter (BS). Both of the beams are coupled into single mode optical fibers (SMF) with lenses (L). One of the fibers is connected to an avalanche photodiode (APD) whilst the other stimulates the rod. Inset shows the pseudo-thermal light source: L is a lens, GD is a rotating ground glass disc, D is a diaphragm. (b) Microscope image (top view) of the fiber taper and a suction micropipette with the constrained rod cell.}
\end{figure}

Methods of cells preparation, electrophysiology recordings, and light coupling were similar to the ones we described previously  \cite{our}. In brief, dark adapted adult male frogs \textit{Xenopus laevis} were used for the experiment. All procedures with the animals were carried out in accordance with IACUC regulations. Shortly after the sacrifice, the retinas were taken from the eyes under IR-light (850 nm) and sliced with a blade to release single rods. The rods were loaded into a  Ringer solution filled chamber of an inverted microscope (Leica, DMI 3000), which was placed in a light tight Faraday cage. The microscope was equipped with an IR LED (920 nm), and a CCD camera (Leica, DFX 380) to allow observation of the preparation without bleaching the rods. A glass micropipette, pulled to the diameter of $\approx$5$\mu m$ and bended by about 45$^\circ$ was used for conventional suction electrode recordings \cite{Bailorsuction}. OS of intact rods were captured in the pipette by gentle air suction, see Fig.1b. The photocurrent pulses were amplified and digitized by a patch-clamp amplifier (Heka, EPC 10), the data were stored by the computer. The experiment was synchronized by an external pulse generator, which triggered data acquisition by the rod and the APD, and conducted at room temperature. Each cell was used for continuous recordings during 90-130 min; 1-2 cells per preparation were used. The consistency  of the response during the experiment was checked prior to acquisition of every point by exciting the saturation amplitude $A_{S}$ with a dim torch \cite{Bailorsuction}.

A tapered single mode optical fiber (Nanonics, working distance 25-30 $\mu$m, spot size 4 $\mu$m), mounted on a motorized stage (Sutter, resolution 0.3 $\mu m$), was used for light delivery to the rod (see Fig.1b). The fiber tip and the rod were carefully positioned at a distance of 25-30 $\mu m$, so that the stimulus were sent along the rod long axis. Such an "axial configuration" provides a nearly optimal match between the light spot and the rod profile, long absorption path in the rod, and an invariance of the rod response to light polarization \cite{our}.

Actual waveforms of the rod photocurrent in response to stimulation by different numbers of impinging photons, produced by the coherent source, are shown in the inset of Fig.2. In addition, a histogram of rod responses to dim coherent light flashes is shown in Fig.6 of supplementary materials. The dependence clearly indicates the quantization of the response, which supports the notion that rods function as photon number resolving detectors \cite{Rieke}, in accordance with the suggested model. The dependencies of the normalized average photocurrent amplitudes, on the average numbers of photons for coherent and pseudo-thermal sources are shown in Fig.2. Amplitudes were normalized by the saturation value $A_{S}$, and the numbers of photons were normalized by $m_{0}$, the value which initiated a response of a half saturation amplitude, which, for the majority of studied cells, was in the range of 550-2,500 photons/pulse. Lines represent numerical calculations of the averages, with the probability distribution given by (5). The dependencies for the majority of investigated cells yielded parameters in the range: $A_{S}\in$[18,26] pA, $\overline{A_{0}}\in$[0.5,0.85] pA, $\sigma_{A}\in$[0.25,0.6] pA, $\sigma_{D}\in$=[0.1-0.2] pA, and $\eta\in$[0.007,0.01]. Note that $\eta$ accounts for an overall efficiency of the photodetection in a given experiment. 

\begin{figure}
\includegraphics[width=8.0cm]{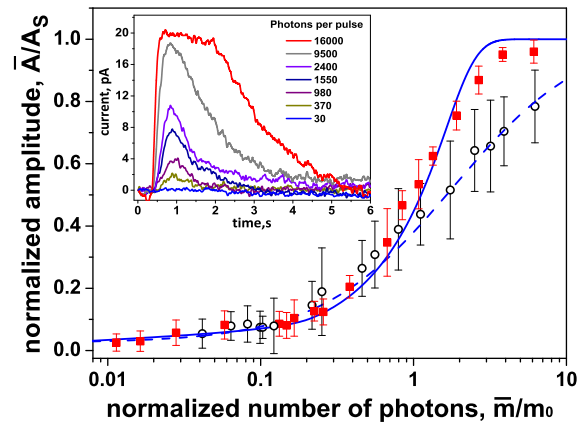}
\caption{(color online) Dependence of the normalized average amplitude of the rod photocurrent on the average number of impinging photons for coherent (red squares, solid line) and pseudo-thermal (black circles, dashed line) sources. Each data point is an average response to 80-100 light pulses. 6 different cells, taken from 6 different animals were used. Lines show numerical calculations, and standard deviation is shown by error bar. Inset shows raw waveforms of the rod photocurrent in response to coherent pulses of different average intensity.}
\end{figure}

The dependencies in Fig.2 demonstrate that saturation of the average amplitude for the coherent source is significantly steeper than for the pseudo-thermal one. A clear explanation for the difference can be found from the experimental histograms of the normalized
amplitudes, which are shown for both of the studied light sources in Fig.3. At relatively low number of impinging photons rod operates in a linear regime, and the statistics of its responses displays the statistical features of the incoming light (Fig.3 a,c). However, when the rod is stimulated by bright flashes, its saturation causes damping of response fluctuations and modifies the light statistics (Fig.3 b,d). For the coherent source photon numbers are well localized around their average value (see (1)), and the majority of response amplitudes to bright flashes are close to the saturation amplitude $A_{S}$ (Fig.3b). In contrast, for the pseudo-thermal source with the same average photon number, the probability of emission of pulses with relatively low photon numbers remains non-negligible (see (1)). Hence the amplitude probability distribution gets significantly broader than in the case of the coherent source (Fig.3b), resulting in smoother saturation of the average response, as shown above (see Fig.2).

Note that the capability of the entire visual system of living human subjects to distinguish between the light sources of different statistics was shown in \cite{Teichmodulation}. Our experiments suggest that responses of single rods are at the basis of this phenomena. Our results also imply that the light source characteristics should be carefully considered for faithful comparison of  experiments on visual transduction in rods.

\begin{figure}
\includegraphics[width=8.0cm]{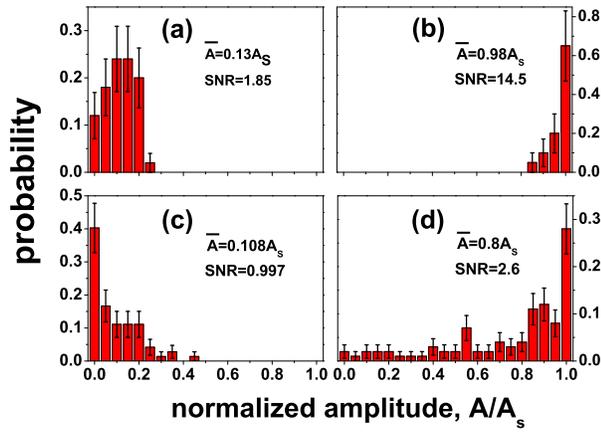}
\caption{(color online) Probability histograms of normalized amplitudes measured for coherent (a,b) and pseudo-thermal (c,d) sources at different values of the average, shown with the corresponding SNR values. Each histogram was acquired for 80-100 light pulses. Standard deviation is shown by error bar.}
\end{figure}

The effect of saturation is evident from the experimental dependence of the SNR on the number of impinging photons, shown along with numerical calculations using (5), in Fig.4a. The raw  data for the measurements of SNR, obtained for 6 different cells, are shown in supplementary materials (see Fig.7). In the linear regime of the rod response, SNR grows linearly for the coherent source, and remains constant for the pseudo-thermal one. At the same time, it demonstrates a steeper growth in the saturation region for both of the sources. Note that saturation of rod should be considered in experiments with non-classical light. In particular, saturation might lead to experimental artifacts, such as observation of sub-shot noise fluctuations ("squeezing") in experiments with classical light sources \cite{calibration}.

\begin{figure}
\includegraphics[width=8.0cm]{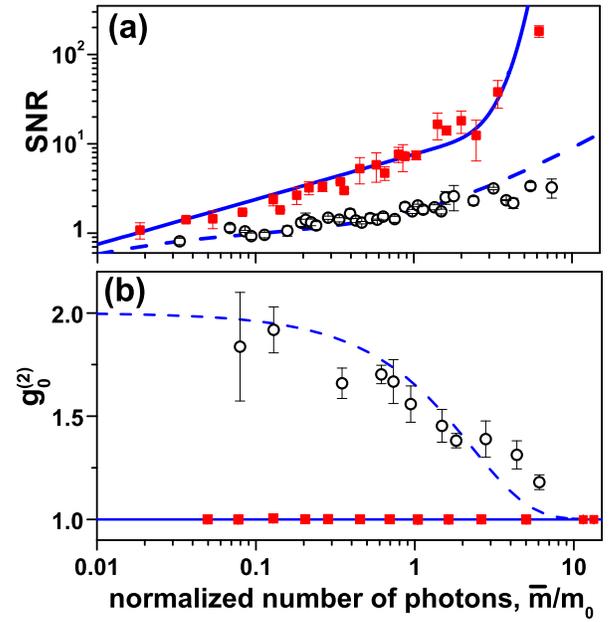}
\caption{(color online) Dependence of the SNR (a) and $g_{0}^{(2)}$ (b) on the average number of impinging photons for coherent (red squares, solid line) and pseudo-thermal source (black circles, dashed line).  Results are averaged  similarly as in Fig.2. Standard deviation is shown by error bar, which can not be seen for the coherent source in (b). Lines show numerical calculations.}
\end{figure}

The experimental dependencies of the correlation function $g_{0}^{(2)}$ on the number of photons for coherent and pseudo-thermal light sources, along with numerical calculations, are shown in Fig.4b. The result for the coherent source yields $g_{0}^{(2)}$=1.01$\pm$0.01 and does not demonstrate any dependence on the number of impinging photons. In contrast, for the pseudo-thermal source, $g_{0}^{(2)}$ decreases with the increasing number of photons. Thus "excess" correlations of the thermal source are gradually neglected due to the saturation of the rod response. As an additional check, in the experiment with the pseudo-thermal source and for the linear regime of the rod response, $g_{\tau}^{(2)}$ was calculated between the photocurrent amplitude and the number of APD counts for consecutive light  pulses. One observed a drop from $g_{0}^{(2)}$=1.94$\pm$0.07, to $g_{\tau}^{(2)}$=0.99$\pm$0.02, which is expected as consecutive light pulses are uncorrelated. Thus it is shown that rods allow adequate measurement of $g_{0}^{(2)}$ similar to conventional man-made detectors, and could be readily used for characterization of quantum light.

In conclusion, being inspired by the ultimate characteristics of rod photoreceptors, we carefully investigated the impact of photon fluctuations of various classical light sources on their response. The experimental results revealed capabilities of isolated rods in measurement of photon statistics. It is of future interest to investigate rods interfaces with nonclassical light, in  particular with correlated two-photon light and intense multi-photon twin-beam states. 
The developed approach can be also used in interfacing rods with realistic sources of intensity fluctuations, for example blinking star lights, when observed through turbulent atmosphere \cite{atmosphere}.

We would like to acknowledge M. Chekhova, M. Tsang, K. Eason, S.-H. Tan, and V. Volkov for stimulating discussions. 
This work was supported by A-STAR Joint Office Council grant and A-STAR Investigatorship grant.


\begin{references}
\bibitem{APD}The number of photons per pulse never exceeded $3\cdot10^{4}$, thus it was perfectly resolved by the APD with a dead time $\tau_{d}\approx35$ nsec.
\end{references}
\end{document}